\pdfoutput=1
\documentclass[graybox]{archivesofdatascience}
\usepackage[T1]{fontenc}
\usepackage{mathptmx}       
\usepackage{helvet}         
\usepackage{courier}        
\usepackage{type1cm}        

\usepackage{graphicx}        
\usepackage{multicol}        
\usepackage[bottom]{footmisc}

\usepackage{amsmath}		
\usepackage{amsfonts}		
\usepackage{latexsym}		
\usepackage{natbib}			

\usepackage{booktabs}		
\usepackage{placeins}		
\usepackage{url}			
\usepackage{hyperref}		

\usepackage[super]{nth}
\usepackage{subcaption}
\usepackage[capitalize]{cleveref}

\newcommand{\creativecommons}{\includegraphics[height=6pt]{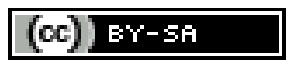}}
\usepackage[activate={true,nocompatibility}, final, tracking=true, kerning=true, spacing=true, factor=1100, stretch=10, shrink=10]{microtype}

\begin{document}

\title*{Enhancing Flood Impact Analysis using \\ Interactive Retrieval of Social Media Images}
\author{Bj{\"o}rn Barz, Kai Schr{\"o}ter, Moritz M{\"u}nch, Bin Yang, Andrea Unger, \\ Doris Dransch, and Joachim Denzler}
\authorrunning{Barz, Schr{\"o}ter, M{\"u}nch, Yang, Unger, Dransch, Denzler}
\institute{Bj{\"o}rn Barz, Joachim Denzler \at Friedrich Schiller University Jena, Computer Vision Group, Ernst Abbe Platz 2, 07743 Jena, Germany \email{{bjoern.barz,joachim.denzler}@uni-jena.de}
\and Kai Schr{\"o}ter, Moritz M{\"u}nch \at Deutsches GeoForschungsZentrum, Sec.\ Hydrology, Telegrafenberg, 14473 Potsdam, Germany \email{{kai.schroeter,moritz.muench}@gfz-potsdam.de}
\and Bin Yang, Andrea Unger, Doris Dransch \at Deutsches GeoForschungsZentrum, Sec.\ Geoinformatics, Telegrafenberg, 14473 Potsdam, Germany \email{{bin.yang,andrea.unger,doris.dransch}@gfz-potsdam.de}}

\maketitle

\index{Barz, B@\emph{Barz, B}}
\index{Schr{\"o}ter, K@\emph{Schr{\"o}ter, K}}
\index{M{\"u}nch, M@\emph{M{\"u}nch, M}}
\index{Yang, B@\emph{Yang, B}}
\index{Unger, A@\emph{Unger, A}}
\index{Dransch, D@\emph{Dransch, D}}
\index{Denzler, J@\emph{Denzler, J}}
\index{Interactive Image Retrieval}
\index{Relevance Feedback}
\index{Flood Impact Analysis}
\index{Machine Learning}

\abstract{The analysis of natural disasters such as floods in a timely manner often suffers from limited data due to a coarse distribution of sensors or sensor failures.
This limitation could be alleviated by leveraging information contained in images of the event posted on social media platforms, so-called ``Volunteered Geographic Information (VGI)''.
To save the analyst from the need to inspect all images posted online manually, we propose to use content-based image retrieval with the possibility of relevance feedback for retrieving only relevant images of the event to be analyzed.
To evaluate this approach, we introduce a new dataset of 3,710 flood images, annotated by domain experts regarding their relevance with respect to three tasks (determining the flooded area, inundation depth, water pollution).
We compare several image features and relevance feedback methods on that dataset, mixed with 97,085 distractor images, and are able to improve the precision among the top 100 retrieval results from 55\% with the baseline retrieval to 87\% after 5 rounds of feedback.}

\section{Introduction}
\label{BARZsec:intro}

The rapid analysis of recent or current natural disasters such as floods is crucial to provide information about impacts as a basis for efficient disaster response and recovery.
With an increasing availability of data and information channels, the identification and exchange of core information and best possible up-to-date information is essential for disaster response \citep{turoff2002past,comfort2004coordination}.
For recovery and improved disaster risk reduction, comprehensive image-based documentations of disaster dynamics help to gain insights and to improve our understanding of system behavior during extreme events.
This knowledge is important to review and adapt flood prevention and protection concepts which are the basis to mitigate adverse consequences from flooding.
However, event analyses often suffer from limited or insufficient data \citep{poser2010volunteered,thieken2016flood}.
For the case of flood mapping, traditional measuring devices as, for instance, water level gauges are expensive and hence only coarsely distributed.
Malfunction and uncertainties of recordings during extreme events are known issues. 

On the other hand, there is usually a large amount of complementary information that could be derived from images posted by volunteers on social media platforms \citep{assumpcao2018citizen}.
Since most modern consumer devices are GPS-enabled and store the geographical location where an image has been taken in its metadata, these images could be used to derive information about the flood at locations where the sensor coverage is insufficient, to substitute failures of measurements, or to complement other pieces of information \citep{schnebele2013improving,fohringer2015social}.
The users of social media platforms posting images about natural disasters can thus be considered as \textit{human sensors} providing so-called \textit{volunteered geographic information (VGI)} \citep{goodchild2007vgi}, from which different types of information can be extracted for being combined with the data obtained from traditional sensors.

These data are potentially useful during all stages of disaster management \citep{poser2010volunteered}:
To prepare for the case of a natural disaster, sufficient data about past events are necessary.
During an event, the rapid availability of social media images would leverage monitoring the extent and intensity of the disaster and the current status of response activities.
For post-disaster recovery, on the other hand, up-to-date damage estimates are required for financial compensation, insurance payouts, and reconstruction planning.

However, the sheer amount of almost 6,000 tweets posted every second on Twitter alone \citep{twitterstats} renders inspecting all of them intractable, even when the set of images is restricted to those within a certain region and time-frame.
Therefore, an automated filter retrieving only those images that are relevant for the analysis is highly desirable.

The notion of \textit{relevance} is usually not fixed but depends on the objective currently pursued by the analysts.
In the case of flood impact analysis, hydrologists might sometimes be interested in determining whether a certain area is flooded or not, which might be difficult to detect based on just a few water level measurements or due to mobile flood-protection walls that may alter the flooding process and expected inundation areas.
However, while VGI images can be of great benefit for this task, just retrieving all images of the flooding is not sufficient in general.
Because at another point of time, the information objective of the analysts might be to determine inundation depth as a key indicator of flooding intensity.
In this regard, a different set of images would be relevant, showing visual clues for inundation depth such as partially flooded traffic-signs or people walking through the water.
Another example is the task of determining the type and degree of water pollution from images, which changes the notion of relevant image features drastically.

These image characteristics are sometimes difficult to verbalize in natural language.
An example image, however, can often capture the search objective much more easily.
Moreover, text-based search always runs the risk of missing relevant images with insufficient textual descriptions.
Thus, we propose an approach based purely on the image content.

\begin{figure}[tbp]
    \includegraphics[width=\linewidth]{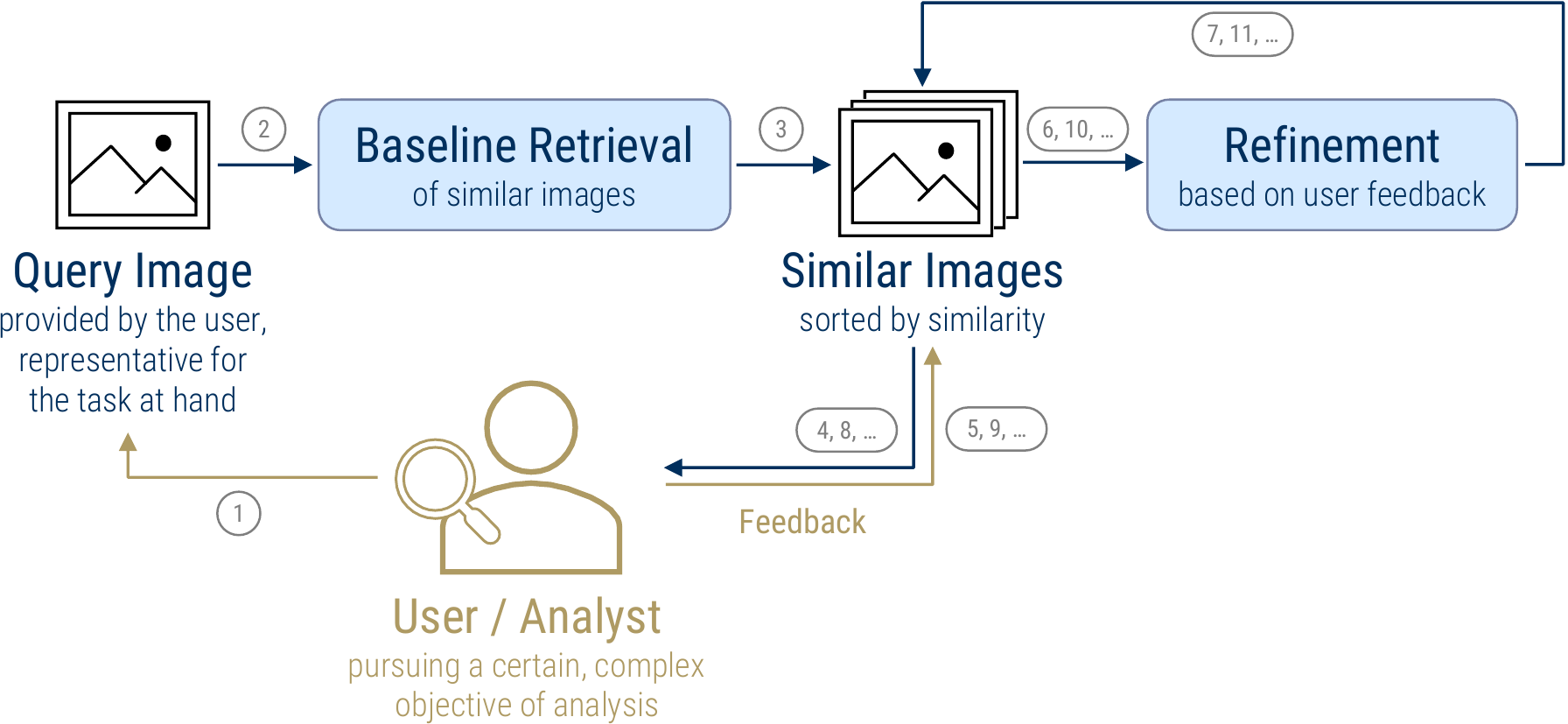}
    \caption{Schematic illustration of our interactive image retrieval process.}
    \label{BARZfig:iir-pipeline}
\end{figure}

Since all the information objectives an analyst might have in mind constitute an open set, it is not possible to train a fixed set of classifiers for distinguishing between relevant and irrelevant samples.
Instead, we propose an \textit{interactive image retrieval} approach to assist the analyst in finding those images that are relevant with respect to the current task.
This procedure is illustrated in \cref{BARZfig:iir-pipeline}:
The user first provides a so-called \textit{query image} that should capture the search objective reasonably well.
The system then extracts image features for this query and compares it with all other images in the database or social media image stream using the Euclidean distance between images in the feature space.
The result is a list of retrieved images, ranked by their proximity to the query.
This procedure is known as \textit{content-based image retrieval} \citep{smeulders2000cbir} and has been an active topic of research since \citeyear{niblack1993qbic} (\citeauthor{niblack1993qbic}).

However, the results of this baseline retrieval will be suboptimal in most cases, since it is based on just a single query image.
Thus, the system enables the user to flag some of the retrieved images as relevant or irrelevant.
This information will then be incorporated to obtain a refined list of retrieval results, which should match the search interest pursued by the user more precisely.
This step can be repeated until the user is satisfied with the result.

In this work, we investigate how to construct an image retrieval pipeline that is suitable for retrieving flood images by comparing several types of features extracted from deep neural networks for the baseline retrieval and various approaches to incorporate relevance feedback.
To enable a quantitative evaluation, we introduce a novel dataset comprising 3,435 images of the European Flood 2013 from Wikimedia Commons plus 275 images showing water pollution from various sources.
All images have been annotated by domain experts with respect to their relevance regarding three pre-defined information objectives.

The remainder of this paper is organized as follows:
We will first briefly review related work on using VGI images for disaster management in \cref{BARZsec:related-work}.
Our novel flood dataset is introduced in \cref{BARZsec:dataset} and various baseline retrieval and relevance feedback methods are described and evaluated in \cref{BARZsec:retrieval}.
\Cref{BARZsec:conclusions} concludes this work and discusses directions for future research.

\section{Related Work}
\label{BARZsec:related-work}

Many approaches for leveraging VGI from social media focus on linguistic patterns \citep[e.g.,][]{ireson2009local}, text-based classification \citep[e.g.,][]{sakaki2010earthquake,yin2012social} and keyword-based filtering \citep[e.g.,][]{vieweg2010microblogging,fohringer2015social}.

Similar to our motivation, \citet{schnebele2013improving} used volunteered data which have been retrieved using the photo, video and news Google search engines for a flood in Memphis (US) in May 2011.
These information have been combined with remote sensing, digital elevation and other data to produce flood extent and flood hazard maps. 

Twitter messages have been used by \citet{brouwer2017probabilistic} to estimate flooding extents.
While this approach uses only Twitter text message contents, it applies a set of keywords to filter relevant tweets.
Geolocation information is derived from location references contained in the tweet.

\citet{fohringer2015social} proposed to derive information about flood events from images posted on Twitter or Flickr and found them to contain ``additional and potentially even exclusive information that is useful for inundation depth mapping''.
Likewise, \citet{rosser2017rapid} retrieve geotagged imagery from Flickr using a defined study area and time window in combination with the keyword ``flood''.
In this approach, only the image location is used to delineate flooded areas in combination with other data sources.
However, both works do not employ any automatic image-based filtering but collect all tweets containing some predefined keywords and then analyze the relevance of all images included in these tweets manually.
On the one hand, this tedious process is prohibitive for rapid flood impact estimation due to the time needed for inspecting all images.
Further, the initial keyword-based filtering involves the risk of missing a large portion of relevant images due to the lack of matching keywords in the text.

We show how these issues can be overcome using computer vision techniques for filtering based on the image content only.

\section{A Dataset for Flood Image Retrieval}
\label{BARZsec:dataset}

A quantitative evaluation of our interactive image retrieval approach demands a sufficient number of both flood and non-flood images.
In addition, we need to know for each image whether it is relevant for a certain task or not.

While obtaining non-flood-related images is rather easy, since any existing dataset such as, for example, the Flickr100k dataset \citep{Philbin07flickr100k} comprising 100,031 images from Flickr could be used for that, finding a sufficient number of images relating to a certain flood event is more difficult.
We used Wikimedia Commons as a source for flood images, since it already provides dedicated categories for major flood events, the images are released under a permissive Creative Commons license, and many of them contain geotags. 

In the following subsections we describe how we collected flood images and annotated them with respect to their relevance regarding three exemplary information objectives.
The dataset, including metadata and annotations, can be obtained at \url{https://github.com/cvjena/eu-flood-dataset}.

\subsection{Collecting Flood Images}
\label{BARZsubsec:image-collection}

The Wikidata project strives towards creating machine-readable representations of all structured information present in Wikipedia.
This information can be queried fully automatically using the SPARQL query language, which allows retrieving a list of all flood events recorded on Wikipedia with an associated Wikimedia Commons category.
We can then use the Wikimedia Commons API to fetch all images and their metadata from those categories automatically.
With a large margin, the highest number of images is available for the Central Europe floods of 2013\footnote{\url{https://commons.wikimedia.org/wiki/?curid=26466898}\\(accessed: July \nth{21}, 2017)}, which comprises 3,855 images in total (as of July 2017).
We hence decided to use this event as a basis for our flood dataset.
After excluding sub-categories that relate exclusively to public transportation during the flood but do not show actual flooding, a total of 3,435 images remain.

However, these images do not show any water pollution, in which we are also interested.
Thus, we added another set of 275 images to the dataset, which we have collected manually from the web by querying image search engines for the names of recent major oil spill events.
To this end, we have again used a list of oil spills provided on Wikipedia\footnote{\url{https://en.wikipedia.org/w/index.php?title=List_of_oil_spills}\\(accessed: May \nth{30}, 2018)}.

\subsection{Relevance Annotations}
\label{BARZsubsec:relevance-annotations}

\begin{figure}[bp]
    \centering
    \includegraphics[width=.86\linewidth]{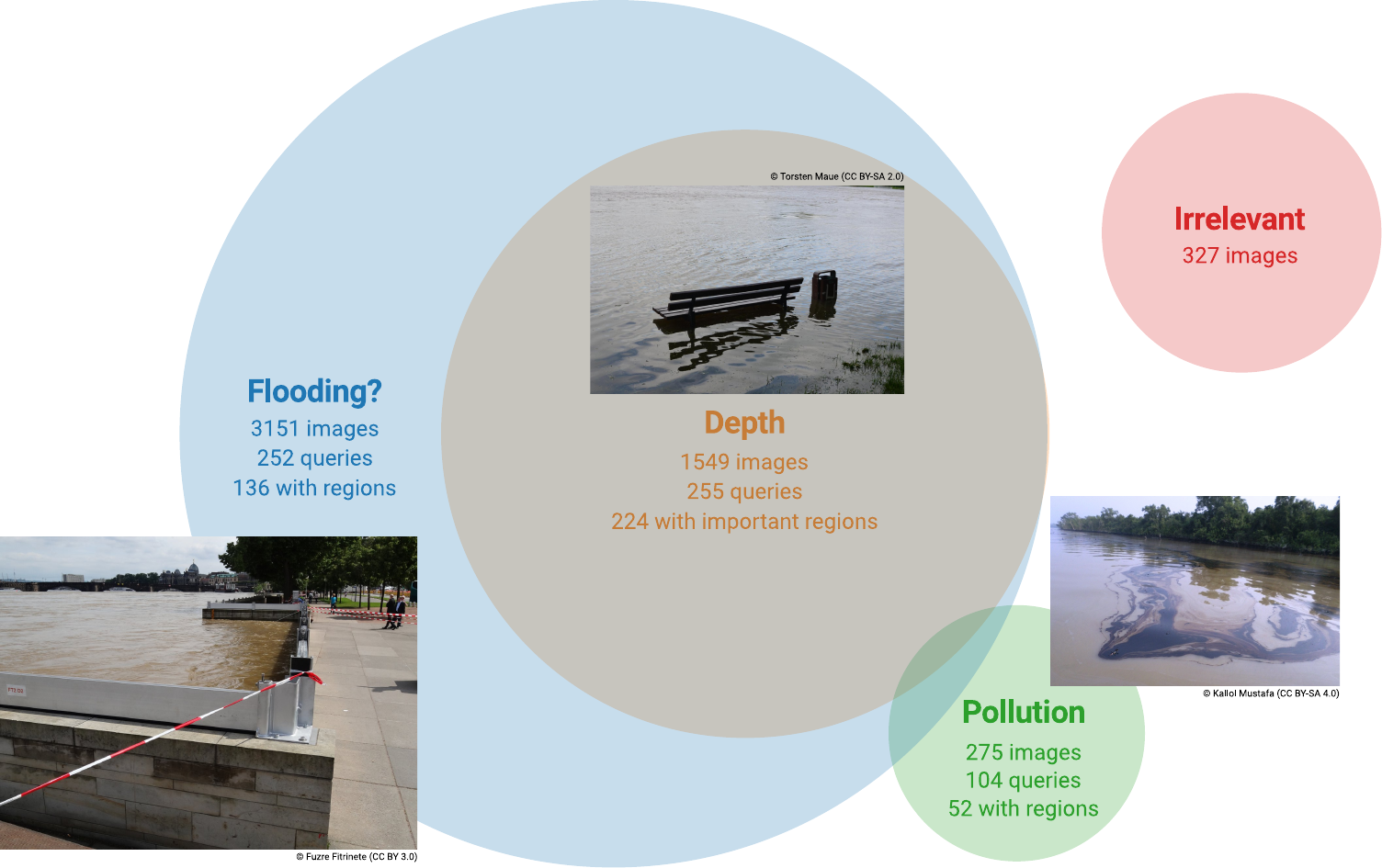}
    \caption{Venn diagram of the sets of images per task in our novel dataset.}
    \label{BARZfig:relevance-statistics}
\end{figure}

For a quantitative evaluation and comparison of several image retrieval methods, we need to simulate the behavior of a user of our proposed interactive image retrieval system.
To enable such a simulation, we have defined a set of three common tasks, which could be pursued by a hydrologist using the system:

\begin{description}
    \item[\textit{Flooded vs. dry}] Does the image help to determine whether a certain area is flooded or not?
    Usually, one would assume flooding of a certain area based on the intersection of the water level height and the elevation of the terrain.
    However, the area might actually be dry due to a flood-protection wall, for example.
    An image considered as relevant would show the boundary between flooded and dry areas.
    Images that do not show any inundation at all are considered not relevant.
    While these could be used to track the spread of the flood at a certain location over time, we only consider the individual relevance of images in this work, ignoring aspects that might become relevant when compared with other images in the dataset.
    
    \item[\textit{Inundation depth}] Is it possible to derive an estimate of the inundation depth from the image due to visual cues such as, for example, traffic signs or other structures with known height?
    If there is no flooding at all, the image is considered as not relevant for inundation depth.
    
    \item[\textit{Water pollution}] Does the image show any pollution substances?
    The focus is on heavy contamination by chemical substances such as oil, for example. 
\end{description}

\noindent
Each image in the dataset has been assigned to one of several domain experts for annotation.
Any image could be relevant for one, multiple, or none of the tasks described above.

\Cref{BARZfig:relevance-statistics} shows the number of images marked as relevant for each task, the overlap of the categories, and an example image for each task.
9.5\% of the images were found not to show any flooding situation despite being associated with the flood in Wikimedia Commons.
We assigned the special label ``irrelevant'' to these images and treat them in the same way as the distractor images from the Flickr100k dataset (see \cref{BARZsec:retrieval}).

Due to limited resources, we only obtained a single annotation for each image.
However, an additional domain expert was asked to assure the quality of random samples from the set of annotations and to select between 100 and 250 ideal query images for each task that reflect the search objective well and could be used as initial query images for our image retrieval approach.

\subsection{Important Image Regions}
\label{BARZsubsec:region-annotations}

\begin{figure}[tbp]
    \begin{subfigure}{.31\linewidth}
        \centering
        \caption{Flooded vs. dry}
        \label{BARZsubfig:areas-flooding}
        \includegraphics[height=2.4cm]{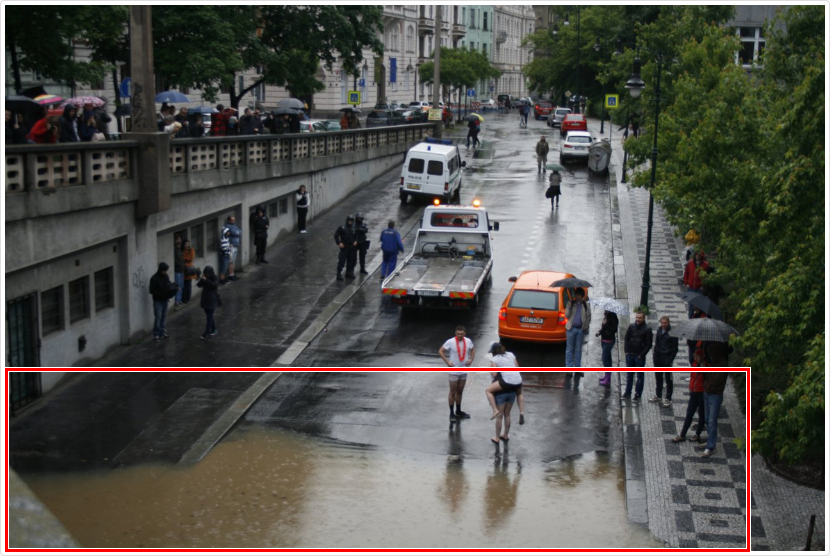}%
        \vspace{-3mm}\flushright\scalebox{.6}{\textcopyright\ Mat\v{e}j Ba\v{t}ha (CC BY-SA 3.0)\;\ }
    \end{subfigure}%
    \hfill%
    \begin{subfigure}{.36\linewidth}
        \centering
        \caption{Inundation depth}
        \label{BARZsubfig:areas-depth}
        \includegraphics[height=2.4cm]{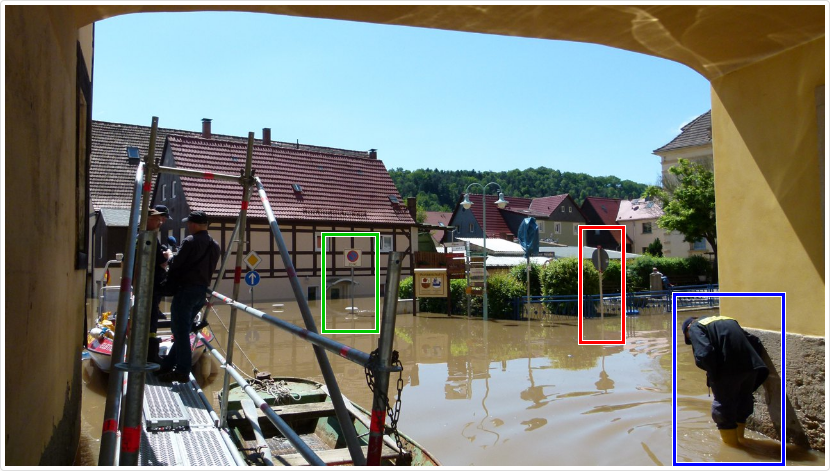}%
        \vspace{-3mm}\flushright\scalebox{.6}{\textcopyright\ Dr. Bernd Gross (CC BY-SA 3.0)\,\ }
    \end{subfigure}%
    \hfill%
    \begin{subfigure}{.31\linewidth}
        \centering
        \caption{Water pollution}
        \label{BARZsubfig:areas-pollution}
        \includegraphics[height=2.4cm]{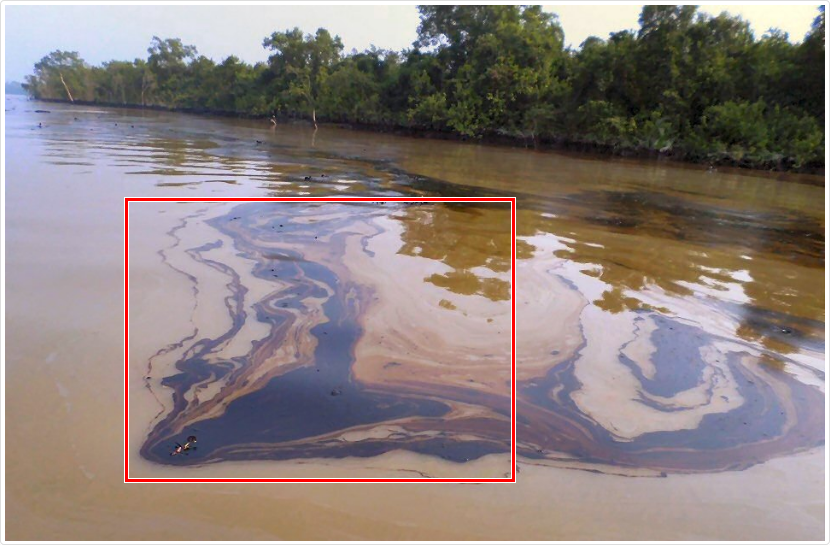}%
        \vspace{-3mm}\flushright\scalebox{.6}{\textcopyright\ Kallol Mustafa (CC BY-SA 4.0)\,\ }
    \end{subfigure}
    \caption{Examples for annotations of important image regions.}
    \label{BARZfig:areas}
\end{figure}

Besides relevance annotations for images as a whole, we have also asked one domain expert to highlight important regions on some of the images selected as queries for each task.
This aims to account for the fact that the relevance of a certain image is often due to a particular small part of the image without which it would not be relevant at all, e.g., partially flooded traffic signs in the case of the inundation depth task.
We also allowed the expert to mark multiple relevant regions per image and to create groups of regions that have to be present together in a single image for being relevant.
Example annotations are shown in \cref{BARZfig:areas}.
We do not make use of these region-level information in our image retrieval system at the moment, but plan to do so in the future.

\section{Interactive Image Retrieval}
\label{BARZsec:retrieval}

In the following, we describe and compare several methods for the two components of our interactive image retrieval pipeline depicted in \cref{BARZfig:iir-pipeline}: constructing a feature space for the baseline retrieval of similar images and incorporating relevance feedback provided by the user.

All methods are evaluated on a combination of our novel flood dataset introduced in \cref{BARZsec:dataset} and images from Flickr100k \citep{Philbin07flickr100k} as distractors.
While Flickr100k comprises a total of 100,031 images, we excluded those tagged with ``river'' or ``water'', since some of them show flooding situations.
After this, 97,085 distractor images remain, which we do not expect to show flooding given their tags.
The set of flood-related images from our novel dataset hence accounts for as few as 4\% of the combined dataset.

We employ the \textit{normalized discounted cumulative gain} \citep{jarvelin2002cumulated} among the top 100 results (NDCG@100) as performance metric, which does not only measure the fraction of relevant images among the top 100 results but considers their order as well, assigning higher weights to earlier positions in the ranking.
For a query $q$ and a ranked list of $n \geq k$ retrieved images with relevance labels $y_i \in \{0,1\}, i=1,\dotsc,n,$ the NDCG@$k$ is defined as:
\begin{equation}
    \label{BARZeq:ndcg}
    \mathrm{NDCG}@k(y_1,\dotsc,y_n \mid q)
    = \left( \sum_{i=1}^{k} \frac{y_i}{\log_2(i+1)} \right)
    / \left( \sum_{i=1}^{\min\{k, |R(q)|\}} \hspace{-3mm} \frac{1}{\log_2(i+1)} \right) ,
\end{equation}
where $R(q)$ denotes the set of all images relevant for the query $q$.
The best NDCG hence is $1.0$ and the worst is $0.0$.
We cap the ranking at $k=100$ since the advantage of our image retrieval system for finding relevant flood images vanishes if the user has to inspect more than 100 results.

In the following, we always report the average NDCG@100 over all 611 query images from our dataset identified as suitable by the domain experts, which are issued as individual queries to the system.
Images from the dataset are considered as relevant with respect to a certain query if they are assigned to the label for which the query image has been selected as ``ideal example''.

\subsection{Baseline Retrieval}
\label{BARZsec:baseline-retrieval}

The main challenge of content-based image retrieval (CBIR) is constructing a feature space where similar images lie close together, so that retrieval can be performed by searching for the nearest neighbors of the query in that space.
The notion of similarity is often fuzzy and depends on the application.
This relation is most often defined as two images either showing the same object, objects of the same class, or being ``visually similar'', which is difficult to formalize.

Traditional CBIR approaches usually consist in detecting invariant keypoints in an image, extracting handcrafted local descriptors from the neighborhood of these keypoints, embedding them in a high-dimensional space, and finally aggregating them into a single global image descriptor.
A good summary of these approaches has been given by \citet{babenko2015aggregating}.

In the past few years, however, such approaches have been outperformed by deep-learning-based image features extracted from a convolutional neural network (CNN) \citep{lecun1989mnistcnn} pre-trained on a classification task.
A CNN typically consists of a sequence of convolution operations with learned filters and non-linear activation functions in-between.
After certain layers, the feature map is sub-sampled using local pooling operations.
The result of the last convolutional layer is hence a low-resolution map of features for different regions of the image.
These local feature vectors are then aggregated by averaging with either uniform or learned weights and fed through a sequence of so-called fully-connected layers, which essentially realize a multiplication of the features with a learned matrix followed by a non-linear activation function.
In classification scenarios, the output of the final layer is interpreted as the logits of a probability distribution over the classes.
The entire network is trained end-to-end using backpropagation \citep{lecun1985backprop}, so that all the intermediate feature representations are learned from data and optimized for the task at hand, where the degree of abstraction from visual to semantic features increases with the depth in the network \citep{zeiler2014visualizing}.

Surprisingly, \citet{babenko2014neural} found these features, which they extracted from the fully-connected layers of a pre-trained network, to also perform competitively for the task of content-based image retrieval.
The traditional CBIR approaches with handcrafted features were finally outperformed using local CNN features extracted from the last convolutional layer \citep{babenko2015aggregating}.
These need to be aggregated into a global image descriptor first, which provides additional leeway for adapting the pre-trained features to the retrieval scenario.
Besides simple average and maximum pooling, a variety of sophisticated pooling functions has been proposed in the past few years.
In this work, we evaluate the following ones:
\begin{description}
    \item[\textit{Average Pooling (avg)}] Uniform average \citep{babenko2015aggregating}.
    
    \item[\textit{Partial Mean Pooling (PMP)}] Averaging over the top 10\% highest activations per channel \citep{zhi2016pmp}. This combines average and maximum pooling.
    
    \item[\textit{Generalized-Mean Pooling (GeM)}] Using the $L^p$-norm of each spatial feature map \citep{radenovic2018gem}, which generalizes between average (for $p=1$) and maximum (for $p\rightarrow\infty$) pooling. We have empirically found $p=2$ to work well on our dataset.
    
    \item[\textit{Adaptive Co-Weighting (adacow)}] Combination of spatial and channel-wise weighting, where the spatial weights are based on the sum of activations at each position and channel weights are determined in a way so that frequently occurring bursty features get a low weight \citep{wang2018adaptive}.
\end{description}

\noindent
In all cases, we extract the local features to be aggregated from the last convolutional layer of the so-called \textit{VGG16} CNN architecture \citep{simonyan2014vgg}, pre-trained\footnote{The pre-trained VGG16 model can be obtained at \url{http://www.robots.ox.ac.uk/~vgg/research/very_deep/} (accessed July \nth{10}, 2019).} for classification on millions of images from the ImageNet dataset \citep{russakovsky2015ilsvrc}.
This is the network architecture that has initially been used by \citet{babenko2014neural} and \citet{babenko2015aggregating} for the first CBIR approaches using neural features and has remained popular until today.
We also evaluate global image features extracted from the first fully-connected (FC) layer of the same CNN, as done by \citet{babenko2014neural}.
This corresponds to a complex aggregation function with learned weights for both feature dimensions and spatial positions.
Regardless of the aggregation function being used, we always $L^2$-normalize the final global image descriptors, which has proven to be beneficial for image retrieval, because the direction of high-dimensional feature vectors often carries more information than their magnitude \citep{jegou2014triangulation,horiguchi2017significance}

Besides the use of pre-trained CNNs for feature extraction, neural networks trained end-to-end specifically for image retrieval have shown superior performance recently.
In this regard, we evaluate the approach of \citet{gordo2016end}, who extended a ResNet-101 architecture \citep{he2016resnet} with R-MAC pooling (sum-pooling over maximum-pooled features from several regions of interest; \citealp{tolias2016rmac}), followed by PCA and $L^2$-normalization.
This network has been trained for image retrieval on a landmarks dataset using a triplet loss, which enforces similar images to be closer together in the feature space than dissimilar ones.
We denote this approach as ``Deep R-MAC'' \footnote{The pre-trained Deep R-MAC model can be obtained at \url{https://github.com/figitaki/deep-retrieval} (accessed July \nth{10}, 2019).}.

\begin{figure}[tbp]
    \resizebox{\linewidth}{!}{%
    \graphicspath{{Graphics/}}%
    \def\svgwidth{20cm}%
    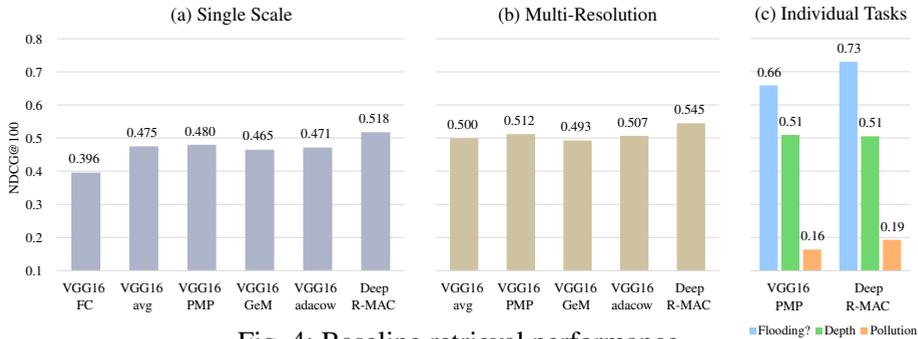%
    }%
    {\phantomsubcaption\label{BARZsubfig:retrieval-results-single}}%
    {\phantomsubcaption\label{BARZsubfig:retrieval-results-multires}}%
    {\phantomsubcaption\label{BARZsubfig:retrieval-results-pertask}}%
    \vspace{-5mm}\caption{Baseline retrieval performance.}%
    \label{BARZfig:retrieval-results}%
\end{figure}

The performance comparison in \cref{BARZsubfig:retrieval-results-single} shows that aggregated convolutional features perform significantly better than features extracted from fully-connected layers, which are presumably already too class-specific.
The choice of the pooling method makes only a slight difference, while PMP performed best.
Features from the Deep R-MAC network fine-tuned for object retrieval provided even better performance than VGG16, resulting in an NDCG@100 of 51.8\% for the simple baseline retrieval.

For these experiments, all images have been resized so that their larger side is 512 pixels wide, except for fully-connected (FC) pooling, which only works with rather small images of size $224 \times 224$ due to the fixed number of learned weights.
Following \citet{gordo2016end} we have also evaluated averaging image descriptors extracted from 3 differently scaled versions of the same image, where we resized the larger side to 550, 800, and 1050 pixels.

The results in \cref{BARZsubfig:retrieval-results-multires} show that the use of multiple resolutions leads to an absolute improvement of NDCG@100 by about 3\%, regardless of the features.

Since the number of images and queries per task in our flood dataset is not balanced (cf.\ \cref{BARZfig:relevance-statistics}), we also report the per-task performance of the two best-performing types of features in \cref{BARZsubfig:retrieval-results-pertask}.
Obviously, finding images relevant for pollution is much more difficult than the other two tasks, which we do not solely attribute to the small number of relevant images, but also to the fact that images of oil films are easily confused with photos of abstract art from Flickr.

\subsection{Relevance Feedback}
\label{BARZsec:relevance-feedback}

Approaches for incorporating relevance feedback into image retrieval can usually be divided into four categories:
\begin{description}
    \item[\textit{Query Point Movement}] The query vector is modified based on the feedback, e.g., by averaging over the features of all images marked as relevant \citep{rocchio1971relevance}. These approaches belong to the oldest ones in information retrieval, but since their use is very limited, we will not address them in this work.
    
    \item[\textit{Probabilistic}] The distribution of the probability that a particular image is relevant given the feedback is estimated. Here, we investigate the simple kernel density estimation (KDE) method proposed by \citet{deselaers2008learning}.
    
    \item[\textit{Classification}] A classifier is trained for distinguishing between relevant and irrelevant images. In this work, we investigate two approaches for this: an Exemplar-LDA classifier \citep{hariharan2012discriminative} and a support vector machine \citep[SVM,][]{cortes1995support}, falling back to a One-Class SVM \citep{scholkopf2001estimating} if only positive feedback is given.
    
    \item[\textit{Metric Learning}] A new metric $d: \mathbb{R}^D \times \mathbb{R}^D \rightarrow \mathbb{R}$ is applied to the $D$-dimensional feature space $\mathbb{R}^D$, minimizing the distance between relevant images and maximizing the distance between relevant and irrelevant ones. Many approaches use a Mahalanobis metric of the form $d_M(x_1, x_2) = (x_1 - x_2)^\top M (x_1 - x_2)$ and learn a positive semi-definite matrix $M \in \mathbb{R}^{D \times D}$. This is equivalent to a linear transformation of the data into a new space where the Euclidean distance corresponds to $d_M$ in the original space.
    
    In this work, we investigate the feature weighting approach of \citet{deselaers2008learning}, the diagonal variant of MMC \citep{xing2003distance}, and information-theoretic metric learning \citep[ITML,][]{davis2007information}.
    The first two approaches learn a diagonal matrix $M$, which corresponds to a weighting of individual features, but use different objectives and optimization algorithms: \citet{deselaers2008learning} minimize the ratio of distances between similar and dissimilar samples using gradient descent, while \citet{xing2003distance} employ a convex optimization objective minimizing the distance of similar samples while keeping dissimilar ones away by at least a fixed radius.
    ITML \citep{davis2007information}, in contrast, learns a full matrix $M$, so that similar pairs are closer than a certain threshold and dissimilar ones are apart by at least another threshold. This is possible despite the high dimensionality of the feature space and limited annotations thanks to regularization towards the Euclidean distance as a prior metric.
    We choose the two thresholds needed for ITML on a per-query basis as follows: All pairs of relevant images should be closer to each other than half the distance between the query and the first irrelevant retrieval result. The distance of all images tagged as irrelevant to any relevant image should be greater than the \nth{95} percentile of the distances between the query and all images in the dataset.
\end{description}

\noindent
Additionally, we investigate combinations of the three metric learning methods with the KDE-based approach of \citet{deselaers2008learning}.

The annotations of our flood dataset allow us to completely simulate the feedback process for a quantitative evaluation:
For all the 611 images denoted as ideal queries, we first perform the baseline retrieval and then mark 10 random images out of the top 100 results either as relevant or irrelevant according to their labels.
This is repeated for a total number of 10 feedback rounds and the retrieval quality is evaluated after each round in terms of the NDCG@100.
To get an impression of the variance of the results, we repeat the entire experiment 10 times with different random sub-samples of 75\% of the dataset.

Based on the findings from the previous section, we use the two best-performing types of features (Deep R-MAC and PMP on the last convolutional layer of VGG16), averaged over multiple image scales.

\begin{figure}[tbp]
    \centering
    \resizebox{\linewidth}{!}{%
        \graphicspath{{Graphics/}}%
        \def\svgwidth{20cm}%
        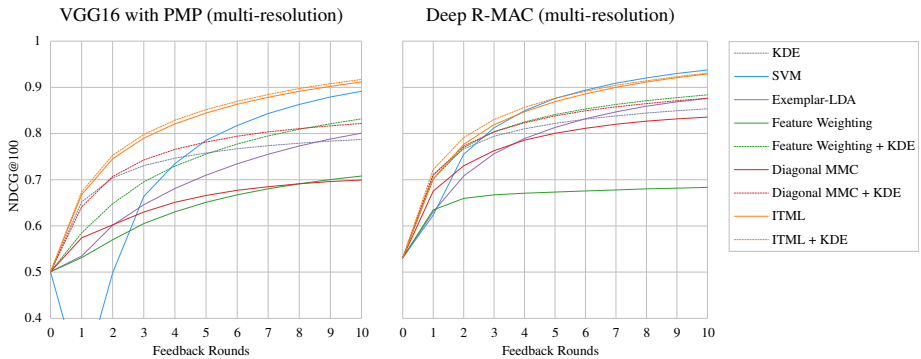%
    }%
    \caption{Comparison of methods for incorporating relevance feedback.}
    \label{BARZfig:feedback-performance}
\end{figure}

The results averaged over the 10 repetitions are shown in \cref{BARZfig:feedback-performance}, where round 0 denotes the performance of the baseline retrieval.
It can be seen that the simple KDE method already performs quite well, especially better than the two feature weighting techniques.
The two classification-based approaches (SVM and Exemplar-LDA) apparently suffer from the limited amount of annotations and behave extremely unstable during the first rounds.
ITML, on the other hand, provides superior performance from the beginning and leads to an NDCG@100 of 86.9\% after 5 rounds of feedback and 92.9\% after 10 rounds.
Though KDE can be added on top of any other method, the benefit when combined with ITML is too marginal to justify the computational overhead.

After 5 rounds of feedback, the SVM-based approach starts to outperform ITML slightly when using Deep R-MAC features.
However, the performance during the first few rounds is of greater importance, since we do not expect most users to regularly spend more than five rounds of feedback for refining the results.
During the early iterations, however, SVM performs worst among all methods.
It might hence be an interesting direction for future work to investigate how ITML and SVM can be combined to improve performance at all stages of the process.

The maximum standard deviation of all methods and all feedback rounds over the 10 repetitions was 1.2\%.
We conducted a paired Student's $t$-test to assess the significance of the differences between the methods in \cref{BARZfig:feedback-performance} at a significance level of 5\%.
At the final feedback round, all differences are significant except that between Exemplar-LDA and Diagonal MMC + KDE when using Deep R-MAC features.
At the first round, all differences are significant for VGG16 features and all besides that between ITML and Feature Weighting + KDE for Deep R-MAC features.
For VGG16 features, ITML and ITML + KDE performed significantly better than all other methods across all rounds.
With the Deep R-MAC features, ITML coincided with Feature Weighting + KDE at round 1 and with SVM at round 4, but was otherwise significantly different from the rest.
SVM started to perform significantly better than the rest from round 7 on.
Besides that, Feature Weighting, Feature Weighting + KDE, and Diagonal MMC, performed significantly different from the rest in at least 9 of 10 rounds.

\section{Conclusions and Future Work}
\label{BARZsec:conclusions}

We have proposed an interactive image retrieval approach with relevance feedback for finding flood images on online image platforms that are relevant for a particular information interest.
To evaluate our approach, we have presented a novel dataset comprising 3,710 flood images annotated with relevance labels regarding three exemplary search objectives and important image regions.

For the baseline retrieval, Deep R-MAC features \citep{gordo2016end} averaged over multiple image scales perform best.
Convolutional features extracted from other networks not fine-tuned for object retrieval can also perform well when aggregated using partial mean pooling \citep{zhi2016pmp}.

Regarding the incorporation of relevance feedback, an SVM-based approach provides the best performance in the long run, but needs a substantial amount of feedback for being useful.
Information-theoretic metric learning \citep{davis2007information}, on the other hand, provides superior performance during the early feedback rounds and remains competitive with SVM later on.
Finally, the simple KDE method of \citet{deselaers2008learning} has turned out to be a quick and decent baseline as well, which is particularly easy to implement and combine with existing frameworks.
Using relevance feedback, the average NDCG@100 can be improved from 55\% yield by the baseline retrieval to 87\% after five rounds and 93\% after ten rounds of feedback, which we expect to be useful for hydrologists to find relevant images quickly.

In the future, we would like to investigate how the selection of important image regions can be integrated as an additional component of the system to improve the relevance of the retrieved images even further, as proposed by \citet{freytag2015interactive}, for example.
It seems also appealing to combine ITML with the SVM-based approach to improve the performance at all stages of the feedback process.
Moreover, it seems promising to apply active learning methods for asking the user for feedback regarding certain actively selected images from which the system expects the most benefit.
Finally, the interactive image retrieval system should be integrated into a visual analytics interface providing data from other sensors as well, enabling a case-study on a more recent flood event with real users.

\begin{acknowledgement}
This work was supported by the German Research Foundation as part of the
priority programme ``Volunteered Geographic Information: Interpretation,
Visualisation and Social Computing'' (SPP 1894, contract number DE 735/11-1).
\end{acknowledgement}

\FloatBarrier
\bibliographystyle{spbasic}
{
\FloatBarrier
\bibliography{Bibliographies/barz}
}

\end{document}